# 2D Ambipolar Vertical Transistors as Control-free Reconfigurable Logic Devices

Zijing Zhao, Shaloo Rakheja, and Wenjuan Zhu

*Department of Electrical and Computer Engineering, University of Illinois at Urbana-Champaign, Urbana, IL 61801, USA*

**Abstract:**

As transistor footprint scales down to sub-10 nm regime, the process development for advancing to further technology nodes has encountered slowdowns. Achieving greater functionality within a single chip requires concurrent development at the device, circuit, and system levels. Reconfigurable transistors possess the capability to transform into both n-type and p-type transistors dynamically during operation. This transistor-level reconfigurability enables field-programmable logic circuits with fewer components compared to conventional circuits. However, the reconfigurability requires additional polarity control gates in the transistor and potentially impairs the gain from a smaller footprint. In this paper, vertical transistors with ambipolar $MoTe_2$ channels are fabricated using the transfer-metal method. The efficient asymmetric electrostatic gating in source and drain contacts gives rise to different Schottky barriers at the two contacts. Consequently, the ambipolar conduction is reduced to unipolar conduction due to different Schottky barrier widths for electrons and holes. The current flow direction determines the preferred carrier type. Temperature-dependent measurements reveal the Schottky barrier-controlled conduction in the vertical transistors and confirm different Schottky barrier widths with and without electrostatic gating. Without the complexity overhead from polarity control gates, control-free vertical reconfigurable transistors promise higher logic density and lower cost in future integrated circuits.

**Introduction:**

As electronic devices approach the feature size limit of manufacturing, reconfigurable field effect transistors become promising for logic applications.[1-5] Reconfigurable transistors can operate as either n-type or p-type, depending on the polarity control signal. Therefore, they can be configured dynamically in run-time.[6-7] Reconfigurable transistors are compatible with CMOS logic and require fewer components in circuits compared to regular transistors, which translates to higher density and lower cost.[6-9] In addition, the ambipolarity of transistors shows functional polymorphism in the same circuit layout, encrypting the circuit function and protecting the actual logic against reverse engineering.[10-12] Non-volatile reconfigurable transistors combine the functionality of memory and logic, leading to a highly efficient in-memory computing scheme.[13] Reconfigurable transistors use polarity control gates to tune Schottky barriers under different configurations and, therefore, do not require source-drain doping.[14-16] The electrostatic doping from polarity control gates provides additional electrons or holes and filters the preferable carrier at the Schottky contact. Despite enabling the reconfigurability of transistors, the additional polarity control gates also impact circuits' performance negatively.[17] These gates may use different voltage levels to operate which adds complexity in power circuitry. The wiring of polarity control gates adds routing penalties in circuits. Moreover, since the reconfiguration needs to switch the voltage on polarity control gates, which adds additional steps in the power up sequence. These drawbacks hinder the integration of reconfigurable transistors into circuits.

Vertical field effect transistors (VFETs) feature vertical channels from semiconductor films. Organic semiconductor films are frequently used thanks to process compatibility.[18-21] Vertical transistors can serve as control-free reconfigurable transistors. The polarity information is implicitly embedded in the current flow direction owing to the asymmetric contact. Therefore, no polarity control gate is required to switch the transistor polarity.[21-22] Since the demonstration of various ultrathin 2D semiconductors, the flexible stacking of 2D materials promotes the development of 2D VFETs. Several vertical transistors use band modulation in tunneling to control the conduction current.[23-25] These transistors can potentially achieve sub-60meV subthreshold swing, promising for low-power logic circuits.[23-25] VFETs based on electrostatic modulation of the Schottky barrier, known as barristor, have been explored in graphene-based 2D/2D and 2D/3D heterostructures.[26-31] The Fermi level of graphene can be tuned dynamically, resulting in modulation of Schottky barrier height at the contact. Vertical barristors allow a higher current level than tunneling-based VFETs and are more promising for high-performance logic circuits.[26-31] Contrary to organic VFETs, the direct modulation of the metal-semiconductor Schottky barrier for both electrons and holes has not been demonstrated in 2D vertical transistors.

Here we propose to implement the control-free reconfigurable logic with 2D ambipolar vertical transistors. The transistor has a simple structure, using ambipolar semiconductor 2D $MoTe_2$ as the vertical channel. Transferred top metal is adopted to allow direct Schottky barrier modulation from the back gate without using graphene heterostructures like barristor. The minimum channel length is as short as 15 nm. The top and bottom contacts are exposed differently to the electrostatic gating from the back gate. Temperature-dependent measurements are conducted to analyze the Schottky barrier-limited conduction. Asymmetric Schottky barriers for different carriers are observed under different current flow directions. Based on this asymmetric conduction, the 2D ambipolar vertical transistor is demonstrated to operate as a control-free reconfigurable transistor. The current flow direction in the channel implicitly controls the transistor polarity. The transistor will switch between n-type and p-type when switching the supply voltage and ground. Therefore, it eliminates the need for an individual control terminal. Control-free reconfigurable transistors promise rapid dynamic reconfigurability for integrated circuits. These results demonstrate the barrier-gating in 2D ambipolar vertical transistors and highlight the potential benefit of "control-free" in reconfigurable transistors, by avoiding the area and power overheads related to polarity control gates.

**Polarity Switching in VFET Based on $MoTe_2$**

The schematic of a typical back gate reconfigurable transistor is shown in Figure 1(a). One (asymmetric) or two (symmetric) polarity control gates may be used, providing additional doping at the contact region, where the Schottky barrier width is tuned to promote unipolar conduction.

The diagram of a control-free vertical reconfigurable transistor is shown in Figure 1(b). The transistor consists of a 2D semiconductor flake sandwiched between two metal pads. The channel is defined as the region of the flake exposed to electrostatic gating. The closest distance between the top and bottom electrodes is equal to the flake thickness. $MoTe_2$ is used here as the channel because its small bandgap allows ambipolar conduction. The $MoTe_2$ channel under the top electrode also overlaps with the back gate. The control-free reconfigurable transistor does not require a polarity control gate. By switching the source and drain terminals, i.e., the current flow direction, this transistor can switch between n-type and p-type polarities. Figure 1(c) gives an optical image of the reconfigurable transistor. The fabrication process starts with metal patterning on a Si wafer with 90 nm $SiO_2$ (Figure 1(d)). Exfoliated $MoTe_2$ flakes with suitable thickness and sizes are picked and transferred on top of the bottom electrode. The top electrode is deposited using a transfer process. It is first deposited and patterned on a separate wafer. A layer of Poly(methyl methacrylate) (PMMA) is spin-coated over the metal at 1500 rpm before a 5 min baking in hexamethyldisilazane (HMDS) vapors. The wafer is then cut into pieces and attached to Polydimethylsiloxane (PDMS) films. After releasing the wafer in water, the PDMS/PMMA/metal stack is dried and transferred onto the $MoTe_2$ flakes. The vertical thickness of the $MoTe_2$ flake in Figure 1(c) is around 15 nm, as measured by atomic force microscopy (AFM). Both top and bottom electrodes have a thickness of 20 nm.

The transfer curves of the reconfigurable transistor measured in different current flow directions are shown in Figure 2(a). When the current flows from the bottom electrode to the top electrode, the transistor shows a higher current when the gate is biased with a positive voltage, corresponding to the n-type behavior, while a p-type behavior is observed when the current flow from the top electrode to the bottom electrode. Both types have comparable current levels. The different behaviors with different current directions originate from asymmetric contacts. The top electrode covers and has good contact with the channel, while the bottom electrode screens the electric field from the back gate and only contacts the channel from the side. As a result, the Schottky barrier near the top electrode can be tuned by electrostatic gating from a back gate voltage, while the Schottky barrier near the bottom electrode is less flexible. As shown in Figure 2(b), when the channel is electron-doped with back gate voltage $V_{BG} > 30$ V, the top electrodes see a much narrower electron Schottky barrier compared to the bottom electrode. When the voltage on the bottom electrode is higher than the voltage on the top electrode ($V_{bot} > V_{top}$), the electrons inject from the top electrode and the holes inject from the bottom electrode. The electron conduction in this case is stronger than the reverse direction when $V_{bot} < V_{top}$. This behavior explains the contrast of electron current between the two transfer curves in Figure 2(a). On the other hand, as shown in Figure 2(c), holes see a narrower barrier at the top electrode when the channel is hole-doped at $V_{BG} < -30$ V. Consequently, hole conduction is more pronounced when $V_{bot} < V_{top}$, as opposed to the reverse scenario where $V_{bot} > V_{top}$. Here the conduction is also contributed by the vertical conduction in the overlapped region of the bottom and top electrodes. Since this conduction is not dependent on the gating, a large intrinsic vertical current will mask the channel conduction and a flat line in the drain current is observed when $V_{BG} < -10$ V and $V_{BG} > 10$ V. Typical symmetric MoTe$_2$ transistor shows an ambipolar behavior. By separating the electrostatic doping sensitivity for the source and drain electrodes, the hole or electron branch in the ambipolar transfer curve is suppressed. The carrier injected in the top electrodes with narrowed Schottky barrier width determines the polarity of the transfer curve. Compared with asymmetric electrostatic gating in the control-free vertical reconfigurable transistor, devices with symmetric electrostatic gating but asymmetric Schottky barrier show diode-like ambipolar conduction for different current flow directions.[32-33] The switching of conventional reconfigurable transistors requires voltages applied on the polarity control gates. In the control-free reconfigurable transistor, polarity switching is achieved by switching the supply voltage and the ground in the circuit. By eliminating the polarity control gate, the control-free reconfigurable transistor reduces both the transistor complexity and the logic overhead to control the polarity.

**Schottky barrier limited conduction in VFET**

Various voltage-dependent measurements are conducted to validate the Schottky barrier limited conduction in the control-free reconfigurable transistors. The transfer curves show a strong dependence on the drain voltage in Figure 3(a). The bottom electrode is grounded, and the top electrode voltage varies between 0.1 V - 0.5 V. The transistor shows an n-type behavior for current flow from top to bottom. The increased voltage drop across the Schottky barrier assists more carriers to tunnel through, leading to an increased conduction current. The contrast between the highest current and minimum conduction current is ~450 at $V_{DS}$ = 0.5 V by measuring the current at $V_{BG}$ = 30 V and $V_{BG}$ = 1.5 V. Note that the minimum conduction current is also contributed by the vertical conduction current in the overlapped region of bottom and top electrodes. Therefore, this current path limits the on/off ratio of the transistor. Figure 3(b) shows the output curves at different back gate voltages measured at 200 K. Voltage sweep is performed on the top electrode, and the bottom electrode is connected to the ground. The transistor is in n-type when $V_{top}$ > 0 V and p-type when $V_{top}$ < 0 V. The on/off state currents of the transistor at n-type and p-type polarities are labeled on the four curve branches. In the right half of the graph, the transistor operates in p-type mode, and the current is higher under $V_{BG}$=-40 V (on-state), and the current is lower under $V_{BG}$=+40 V (off-state). Similarly, the left half of the graph shows the comparison of on/off state currents in the n-type mode. When $V_{BG}$ = 0 V, no electrostatic gating is present in the channel. The current consists of both lateral and vertical conductions, and they are limited by the unnarrowed Schottky barriers, showing strong dependence on the voltage between the top and bottom electrodes. We note that the vertical conduction in the overlapped region is independent of back gate voltages. The off-sate currents in the n-type and p-type configurations are close to the leakage current at $V_{BG}$ = 0 V, which confirms that the width of the Schottky barrier at the off-state is closer to that of the Schottky barrier without electrostatic gating. This confirms that the strength of electrostatic gating constitutes the contrast between the on and off states of the control-free reconfigurable transistors.

The temperature dependence of the transfer and output curves helps us understand the Schottky barriers in control-free vertical reconfigurable transistors. The transfer curves of the transistor at n-type and p-type modes are presented in Figure 4(a) and (b) respectively. The off-state current shows a stronger temperature dependence compared to the on-state current. This contrast suggests differences in the Schottky barrier width at the on- and off-states. Figure 4(c) shows the output curve in the log scale, with the p-type mode on the right half and n-type mode on the left half of the graph. The back gate voltage is applied at +40 V and -40 V accordingly to highlight the on-state current. The output curves show a clear temperature dependence at the small $V_{DS}$ regime, confirming that the conduction is limited by Schottky barriers, where thermionic emission and thermal-assisted tunneling are the dominant conduction mechanisms. The Schottky barrier heights are extracted for both n-type and p-type modes and compared with the barrier height without electrostatic gating. The extraction process and results are shown in Figure 4(d). Note that the extracted barrier height is not in flat band condition and the results are effective barrier heights. $|V_{DS}| = 0.15V$ is chosen to limit the barrier-lowering effect introduced by $V_{DS}$. The conduction in the channel is modeled with the conduction across a Schottky barrier, represented by the following equation[33]

$$I_{DS} \propto T^2 \exp\left(\frac{-q\phi_b}{kT}\right)\left(-\exp\left(\frac{qV_c}{nkT}\right) + 1\right), \tag{1}$$

where $V_c$ refers to the voltage across the Schottky barrier; $n$ is the ideality factor, and $\phi_b$ is the Schottky barrier height; $q$ is the electron charge; $T$ refers to the absolute temperature in kelvin, and $k$ is the Boltzmann constant. By plotting $I_{DS}/T^2$ on a logarithmic scale versus $1/T$, the barrier height can be extracted from the slope of curves near room temperature. We obtain 0.18 eV effective barrier heights when the gate voltage is at +40 V and -40 V. A 0.37 eV effective barrier height is obtained for the conduction without electrostatic doping when the gate voltage is 0 V. The effective barrier height is extracted assuming the channel current is due to thermally activated injection 'over' (not through) a barrier only. When the gate voltage is equal to flat band voltage, the extracted effective barrier height corresponds to the actual barrier height of the contact. When the gate voltage is at +40 V, the electron Schottky barrier is narrowed and thermal assistant tunneling current contributes to the conduction, therefore, the conduction shows a weaker temperature dependence and a smaller effective Schottky barrier is obtained. Similarly, when the gate voltage is at -40 V, the hole Schottky barrier is narrowed and a smaller hole effective Schottky barrier is obtained. The modulation range of barrier height here is comparable with barristors,[26] indicating the transferred top metal has a weak Fermi level pinning effect. The extracted barrier height confirms that high electrostatic doping at the on-state narrows the Schottky barrier width and enables more carriers to tunnel through the contact barrier in the transistor. Without electrostatic gating, the Schottky barrier width remains the same.

**Conclusion:**

In conclusion, we demonstrate a 2D vertical transistor with ambipolar $MoTe_2$ as the channel. Temperature-dependent measurements are conducted to analyze the effective Schottky barrier height under different bias conductions. We confirm that the current conduction is limited by the Schottky barrier. The electrostatic gating from the back gate differentiates the electrons and holes injected from the top and bottom electrodes. Efficient modulation of the metal-semiconductor Schottky barrier is enabled by a transferred top electrode. Consequently, carriers injected from the top electrodes encounter a significantly reduced effective Schottky barrier height. In contrast, the bottom electrodes are not gated, resulting in carriers experiencing a larger effective Schottky barrier. The ambipolar conduction of $MoTe_2$ is reduced to unipolar n-type and p-type behavior under different current flow directions. Therefore, when the current flow direction is switched, these transistors can switch polarity without the assistance of polarity control gates. Compared with conventional reconfigurable transistors, the 2D ambipolar vertical transistor avoids the integration difficulties of polarity control gates and promises reconfigurable logic with high density and low cost.

**ACKNOWLEDGMENT**

The authors would like to thank the support from Semiconductor Research Corporation (SRC) under grant SRC 2021-LM-3042.

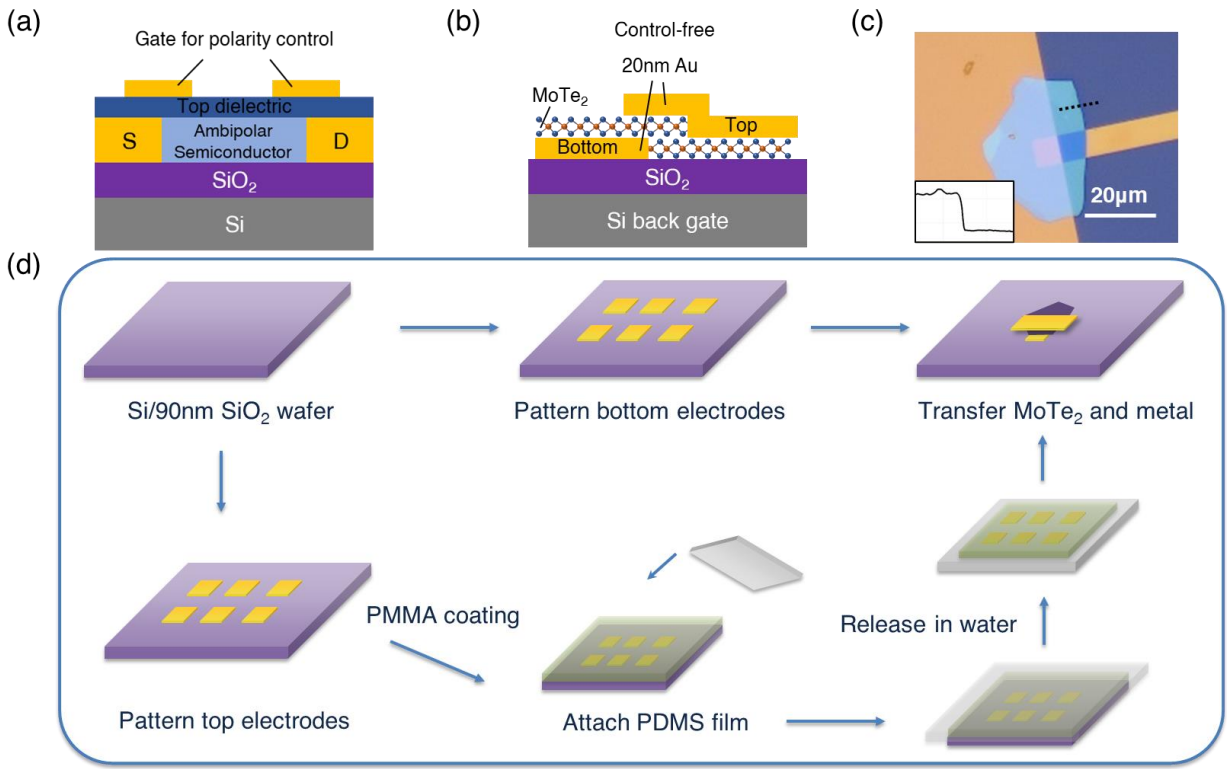

Figure 1 (a) Diagram of a typical reconfigurable transistor. One or two top gates may be used for polarity control. (b) Diagram of a control-free vertical reconfigurable transistor. The transistor has no programming gate and has asymmetric contacts between the source and drain. (c) Optical image and AFM mapping of a typical device. The flake used in the transistor is 14.7nm thick. (d) Fabrication flow of the device. First, MoTe$_2$ flakes are transferred on the prefabricated metal electrode, and the top Au is subsequently transferred on top of the MoTe$_2$/Au stack.

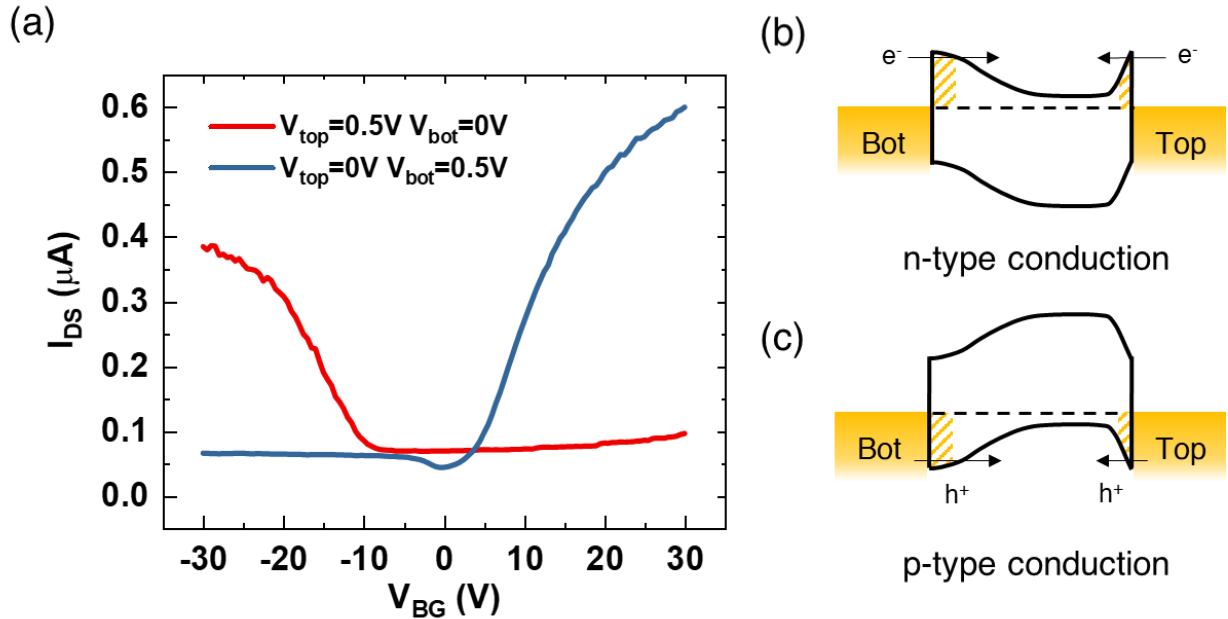

Figure 2. (a) Transfer curves of the VFET with current flowing from top to bottom and bottom to top show p-type and n-type polarities respectively. (c) The conduction of electrons and holes is limited by the Schottky barrier at the contacts under different current flow directions and back gate voltages. For n-type conduction where $V_{BG} > 30$ V, the electrons conduction from the top electrodes sees a narrower barrier due to the electrostatic doping below the contact. Therefore, the electron current is higher when flows from the bottom electrode to the top electrode. The electrostatic gating also narrows the Schottky barrier width for holes in p-type conduction when $V_{BG} < -30$ V. Therefore, the hole current is higher when flowing from the top electrode to the bottom electrode. Depending on the current flow direction, the carrier injected from the top electrodes always sees a narrower barrier than from the bottom electrode, which leads to unipolar transport with switchable polarity.

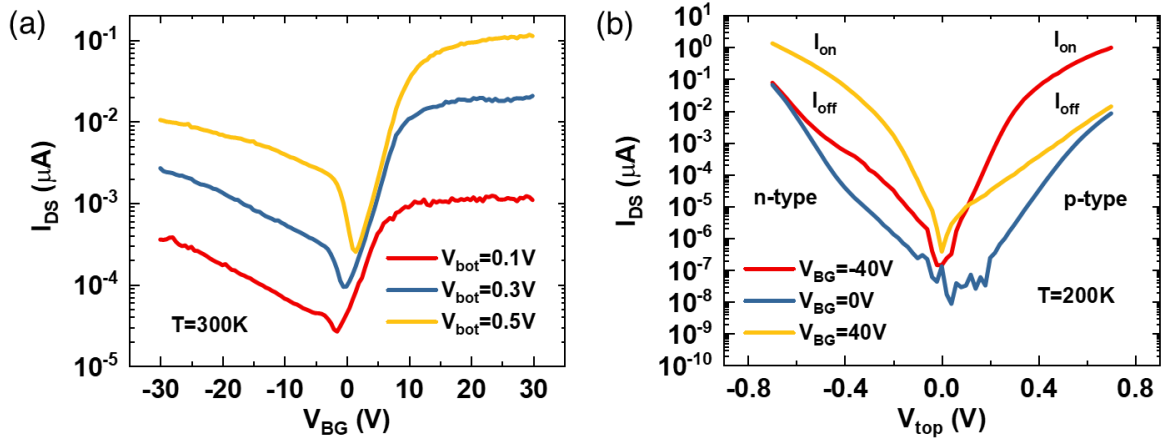

Figure 3. (a) The transfer curves measured at different source-drain voltages. By applying a voltage at the bottom electrode and grounding the top electrode, an n-type behavior is achieved. Increasing source-drain voltage enhances the conduction by lowering the barrier at the injection side. (b) The output curves were measured for n-type and p-type states at different back gate voltages at 200K with the bottom electrode grounded. The on/off conditions refer to $V_{BG}$ = 40 V and $V_{BG}$ = -40 V. Four on/off currents for n-type and p-type polarities are labeled. Both the on and off currents are strongly dependent on the $V_{DS}$. The bottom electrode is grounded during the measurement. The transistor behaves p-type when $V_{top}$ > 0 V and n-type when $V_{top}$ < 0 V respectively. The gap between $I_{on}$ and $I_{off}$ indicates the on/off ratio.

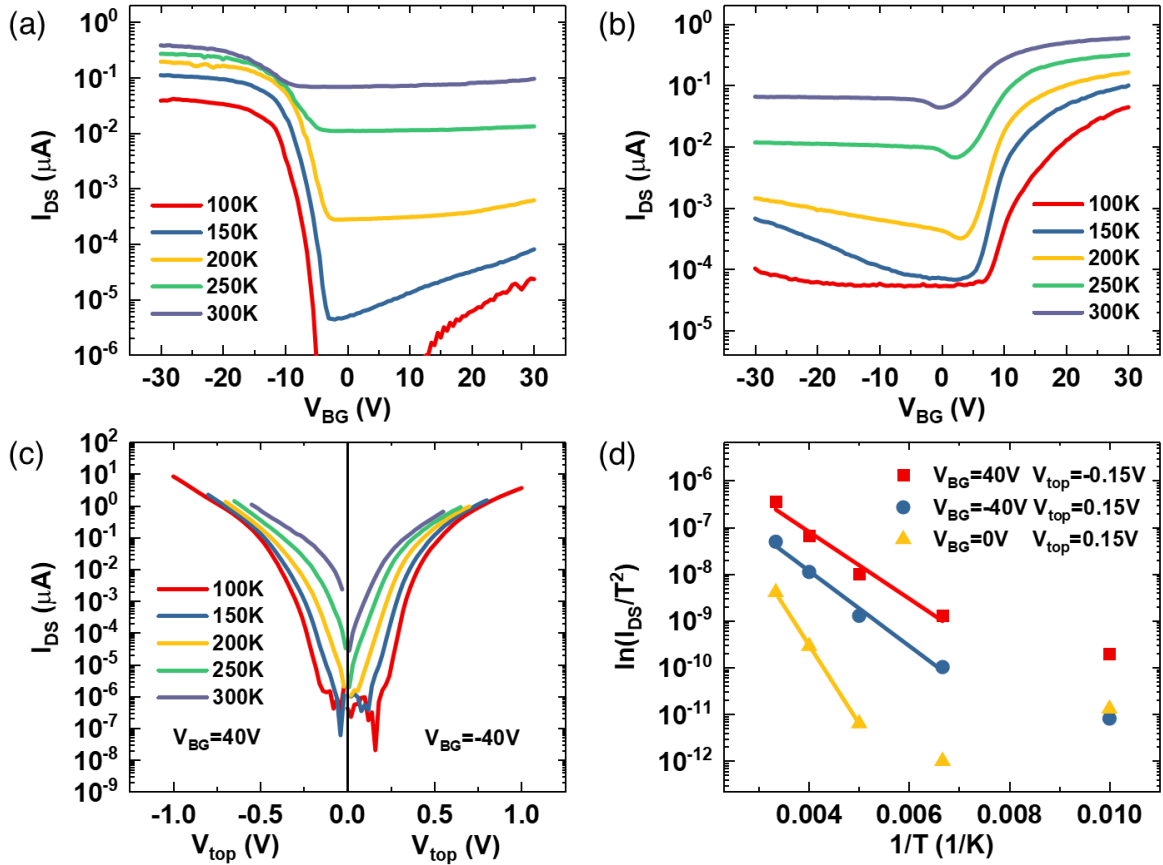

Figure 4. (a) and (b) show the temperature dependences of the transfer curves of n-type and p-type transistors. The off-state current shows a stronger temperature dependence, indicating a larger Schottky barrier in the conduction. The curves are obtained at $|V_{DS}| = 0.5$V. (c) The temperature dependence of on-state current for n-type and p-type polarities. (d) Arrhenius-type plots constructed at different gate voltages. The effective Schottky barrier heights are extracted from the slope near room temperature.